# Quantification of cervical elasticity during pregnancy based on transvaginal ultrasound imaging and stress measurement

Peng Hu[1,†], Peinan Zhao[2,†], Yuan Qu[3], Konstantin Maslov[1], Jessica Chubiz[2], Methodius G. Tuuli[4], Molly J. Stout[5,*], Lihong V. Wang[1,*]

[1]Caltech Optical Imaging Laboratory, Andrew and Peggy Cherng Department of Medical Engineering, Department of Electrical Engineering, California Institute of Technology, Pasadena, CA, USA
[2]Department of Obstetrics and Gynecology, Washington University in St. Louis, St. Louis, MO, USA
[3]Department of Biomedical Engineering, Washington University in St. Louis, St. Louis, MO, USA
[4]Department of Obstetrics and Gynecology, Brown University, Providence, RI, USA
[5]Department of Obstetrics and Gynecology, University of Michigan, Ann Arbor, MI, USA

[†]These authors contributed equally.

*Correspondence should be addressed to Molly J. Stout (mjstout@med.umich.edu) and Lihong V. Wang (LVW@caltech.edu).

**Short title:** Stress-strain cervical ultrasound elastography

**Contribution**

What are the novel findings of this work?

We develop a safe-in-pregnancy, operator-independent, and quantitative cervical elastography system based on transvaginal ultrasound imaging and stress measurement. The system shows high accuracy in phantom experiments and captures the expected decrease in cervical stiffness over pregnancy in a pilot population of pregnant patients.

What are the clinical implications of this work?

The system is safe, robust, and applicable to different ultrasound machines with minor software updates. It can be tested to define normal and abnormal cervical softening patterns in pregnancy for larger populations, which facilitates insights into cervical-elasticity-related diseases such as preterm birth with multiple potential clinical uses.

**ABSTRACT**

**Objective**: Strain elastography and shear wave elastography are two commonly used methods to quantify cervical elasticity; however, they have limitations. Strain elastography is effective in showing tissue elasticity distribution in a single image, but the absence of stress information causes difficulty in comparing the results acquired from different imaging sessions. Shear wave elastography is effective in measuring shear wave speed (an intrinsic tissue property correlated with elasticity) in relatively homogeneous tissue, such as in the liver. However, for inhomogeneous tissue in the cervix, the shear wave speed measurement is less robust. To overcome these limitations, we develop a quantitative cervical elastography system by adding a stress sensor to an ultrasound imaging system.

**Methods**: In an imaging session for quantitative cervical elastography, we use the transvaginal ultrasound imaging system to record B-mode images of the cervix showing its deformation and use the stress sensor to record the probe-surface stress simultaneously. We develop a correlation-based automatic feature tracking algorithm to quantify the deformation, from which the strain is quantified. After each imaging session, we calibrate the stress sensor and transform its measurement to true stress. Applying a linear regression to the stress and strain, we obtain an approximation of the cervical Young's modulus.

**Results**: We validate the accuracy and robustness of this elastography system using phantom experiments. Applying this system to pregnant participants, we observe significant softening of the cervix during pregnancy ($p$-value < 0.001) with the cervical Young's modulus decreasing 3.95% per week. We estimate that geometric mean values of cervical Young's moduli during the first (11 to 13 weeks), second, and third trimesters are 13.07 kPa, 7.59 kPa, and 4.40 kPa, respectively.

**Conclusion**: Due to its simplicity, safety, and ability to provide longitudinal measurements and comparisons between examiners, this quantitative cervical elastography system has the potential for large-scale scientific and clinical applications.

**INTRODUCTION**

Preterm birth (delivery < 37 weeks of gestation) is the most significant cause of neonatal morbidity and mortality[1]. Due to the high risks of prematurity to a neonate, maternal hospitalization in patients with preterm labor symptoms accounts for 1/3 of antepartum hospitalizations. Yet, more than half of women hospitalized with threatened preterm labor go on to deliver at term[2]. For better management, there is a demanding need to develop tools to predict premature delivery more accurately.

Cervical ripening is the process by which the cervix undergoes changes to soften, shorten, and dilate to allow passage of the fetus in delivery. Although manual palpation (Bishop score) has been used to estimate cervical softening, its predictive utility for preterm birth is limited[3]. A quantitative method to document cervical softening before detectable dilation could significantly advance the field of preterm birth prediction.

Ultrasound elastography is commonly used to quantify tissue elasticity due to its wide availability and relatively low cost[4]. It includes stress-strain elastography and shear wave elastography[4–11], which measure tissue Young's modulus and shear wave speed, respectively. As a simplification of stress-strain elastography, strain elastography[4,6,7,10,11] has been widely applied to the breast, prostate, liver, and thyroid where regional tissue-elasticity differences are clinically useful[12–17]. However, without stress information, strain elastography does not support comparisons between different imaging sessions.

Shear wave elastography[4,5,8,9] is also used clinically to measure shear wave speed in the breast[5,18–20], prostate[5,21,22], liver[23–27], and thyroid[28–30], and the measurement allows for longitudinal analysis. Although it has been applied to the cervix[31–39], there are some limitations. First, the homogeneous-tissue assumption for shear wave elastography is violated in the cervix by cystic areas, blood vessels, endocervical canal, and layers of collagen with varying alignments[38,40]. Second, tissue excitation changes that shear waves potentially elicit[41,42] make shear-wave-free ultrasound elastography safer during pregnancy. Third, operator and patient-dependent factors[38,43,44], such as the stress applied to the cervix, may affect the shear wave speed.

To address the above limitations, we develop a protocol based on transvaginal ultrasound that is safe in pregnancy, operator-independent, and quantitative, using stress-strain elastography to capture cervical tissue elasticity.

**METHODS**

**Pregnant participant inclusion**

Pregnant participants in this study were from a prospective, longitudinal cohort study[45] performed at Washington University in St. Louis Medical Center between January 2017 and January 2020. These participants were asked to visit from their first trimester to delivery, with gestational ages from 11 to 38 weeks. For this research, we include those (22 participants) with three or more imaging visits that meet the measurement quality requirement for this study and the last visit greater than 34-week gestational age. Characteristics of these participants are shown in **Supplementary Table 1**. Approval of all ethical and experimental procedures and protocols was granted by the Institutional Review Board of the Washington University in St. Louis. All participants provided written informed consent for the collection and use of clinical, imaging, and questionnaire data.

**Pregnant participant cervical imaging**

During a quantitative cervical elastography imaging session, a trained obstetric sonographer obtains a sagittal view of the cervix which shows the cervical stroma and cervical canal using the Hitachi Noblus ultrasound system with a transvaginal ultrasound probe. The cervix should occupy no less than 2/3 of the imaging depth and be located at the center of the image. Stress is gently applied and released in 2 to 4 cycles (4 seconds) compressing and releasing the tissue along the probe-axis direction, while the B-mode images are continuously recorded to a video (120 frames, 30-Hz frame rate) and the stress is measured by a stress sensor. At least three videos are acquired for each participant in this research. Stress sensor measurements are calibrated after each imaging session of a participant to obtain true stresses, and strains are quantified from frames in each B-mode video. A linear regression is applied to the stresses and strains to estimate Young's modulus of the cervix. The geometric mean value of multiple estimated Young's moduli in an imaging session of a participant is used to represent the Young's modulus of the session. To facilitate the operation, we developed a graphical user interface (GUI) for participant imaging (**Supplementary Note 1**).

**Stress measurement**

We customize a stress sensor (FlexiForce A101) and mount it to the ultrasound probe for stress measurement. The currently used stress sensor can be damaged if directly used in an aqueous environment. To make it waterproof, we embed it into a layer of silicone sealant, which also stabilizes the contact for robust measurement. Before each measurement, both the ultrasound probe and stress sensor are disinfected; then they are dried for stable contact. Next, the ultrasound probe is connected to the ultrasound machine, and the two wires soldered to the stress sensor are connected to a microcontroller board. The stress sensor sampling rate is set as 80 Hz, higher than the frame rate (30 Hz) of the B-mode videos from the ultrasound system. Then the stress sensor is mounted onto the tip of the probe and fixed by two rubber bands. After sensor mounting, a syringe is used to inject ultrasound gel into the space between the silicone sealant and the probe. The stress sensor covers a relatively small area of the detection surface of the ultrasound probe. The ultrasound transmits through the silicone sealant in other parts of the detection surface and the ultrasound image is minorly affected by the stress sensor.

**Stress calibration**

We designed a calibration system using a load cell, a load cell amplifier, and a microcontroller board and the system is used to calibrate the stress sensor after each imaging session. The load cell (maximum weight of 780 g) is calibrated with standard weights right after initial setup and in the following weekly maintenance. For stress sensor calibration, to stabilize the contact, a layer of rubber is placed between the load cell and the stress sensor mounted on the probe. Multiple press-release cycles are applied to the probe along its axial direction so that the stress sensor responses cover the whole range of measurements for cervical tissue. Measurements from the stress sensor and load cell during these cycles are used for calibration. For an imaging session, we denote the calibration function as $f_{\text{cal}}$, which maps the true stress to the stress sensor measurement (**Supplementary Note 2**).

**Strain quantification**

We propose an algorithm to quantify cervical strain from all the frames in each B-mode video. In this research, we analyze the strain only along the ultrasound probe axis. First, we quantify the cervical deformation between every pair of consecutive frames in the video through a correlation-based method and calculate the deformation between two arbitrary frames through recursion (**Supplementary Note 3**). Next, we correct the accumulative error in the recursion based on the periodicity of the deformation (**Supplementary Note 4**). Then we quantify the logarithmic strain of the cervical tissue from the cervical deformation (**Supplementary Note 5**).

**Stress-strain regression**

The measurement from a single stress sensor does not have enough information to determine the stress distribution along the probe axis. In this research, we use the stress sensor measurement to approximate the average stress along the probe axis. We denote the stress sensor measurement corresponding to the $l$-th frame in the B-mode video as $\tilde{\sigma}_l$, which is calibrated to the true stress $f_{cal}^{-1}(\tilde{\sigma}_l)$, and the cervical strain between the first and the $l$-th frame as $\epsilon_l, l = 1, 2, \ldots, L$. Applying a linear regression to the stresses and strains, we obtain

$$f_{cal}^{-1}(\tilde{\sigma}_l) \approx E\epsilon_l + b, l = 1, 2, \ldots, L. \qquad (1)$$

Here, $E$ and $b$ denote the estimated Young's modulus and the intercept, respectively.

**Measurement quality requirement**

During an imaging session, the Pearson correlation coefficient (PCC) between the cervical strains and stress sensor measurements is calculated (denoted as $r'_{ss}$) after each B-mode video is acquired. These videos, cervical strains, stress sensor measurements, and the values of $r'_{ss}$ are used to guide the sonographer to maintain or improve the measurement quality. After an imaging session, we use two parameters to estimate the measurement quality. The PCC between the stress sensor measurements and load cell measurements in a calibration, denoted as $r_{cal}$, is used to estimate sensor performance. To reject errors caused by stress sensor malfunction, we analyze only measurements with $r_{cal} \geq 0.85$. After stress sensor calibration and strain quantification, the PCC between the stresses and strains is calculated, denoted as $r_{ss}$. To reduce errors caused by imperfect operations, only measurements with $r_{ss} \geq 0.8$ are used to estimate Young's moduli. All measurements with $r_{cal} < 0.85$ or $r_{ss} < 0.8$ suggested high noise in the stress or strain measurement and were excluded.

**System accuracy, repeatability, and reproducibility**

We tested the accuracy, repeatability, and reproducibility of the proposed system in phantom and participant experiments. In the phantom experiments, we prepared four gelatin phantoms with concentrations 70 g/L, 90 g/L, 110 g/L, and 130 g/L, respectively. Two operators used two stress sensor replicas, respectively, to perform the quantitative elastography on each phantom 10 times. One of the operators also used the standard quasi-static compression method to measure the Young's modulus of each phantom 10 times. All measurements from both operators were then compared to evaluate the agreements between the two methods (accuracy), between repetitions (repeatability), and between operators (reproducibility). In the participant experiments, two sonographers performed the quantitative elastography on a pilot cohort of 19 participants. Each participant was measured by two sonographers using the quantitative cervical elastography system at least three times during the visit. The measurements were then compared to evaluate the repeatability and reproducibility of the system.

**Statistical analysis**

The PCCs were used to evaluate the linear correlation between the stress sensor measurement and load cell measurement, and that between the strain and true stress. The two-way analysis of variance (ANOVA)[46] was used to test the difference between the proposed method and the standard quasi-static compression method (**Supplementary Note 6**) in measuring elasticity. The coefficient of variation (CV) was used to quantify the repeatability of multiple measurements performed by an operator (a sonographer) on a phantom (participant). The Bland-Altman plot[47] for reproducibility test was used to analyze the agreement between sonographers' measurements. The linear mixed model was used to analyze the decreasing trend of Young's moduli during pregnancy. The gestational ages of the measurements were considered as fixed longitudinal effect to Young's moduli. The individual difference was considered as a random intercept/slope effect to describe the inter-patient variability. A test with *p*-value < 0.05 was considered statistically significant. All analyses were performed using R statistical software v4.1.2 (R Core Team 2021).

RESULTS

**Quantitative cervical elastography system**

The quantitative cervical elastography system, as summarized in **Figure 1a**, is based on three components: a transvaginal ultrasound imaging system, a stress measurement system, and a stress calibration system. Two transvaginal B-mode images of a cervix (posterior boundary marked as a red-dashed curve) are shown in **Figure 1b1** and **b2**, respectively, from which the cervix deformation is observed. A series of quantified strains are shown in **Figure c**. The stress measurement system is visualized in **Figure 1d1-d2** and a series of measured stresses are plotted in **Figure 1e**. The calibration system is depicted in **Figure 1f1-f2** with a calibration function shown in **Figure 1g**, which inversely transforms the stress measurements to true stresses (**Figure 1h**) and allows for the final linear regression shown in **Figure 1i**.

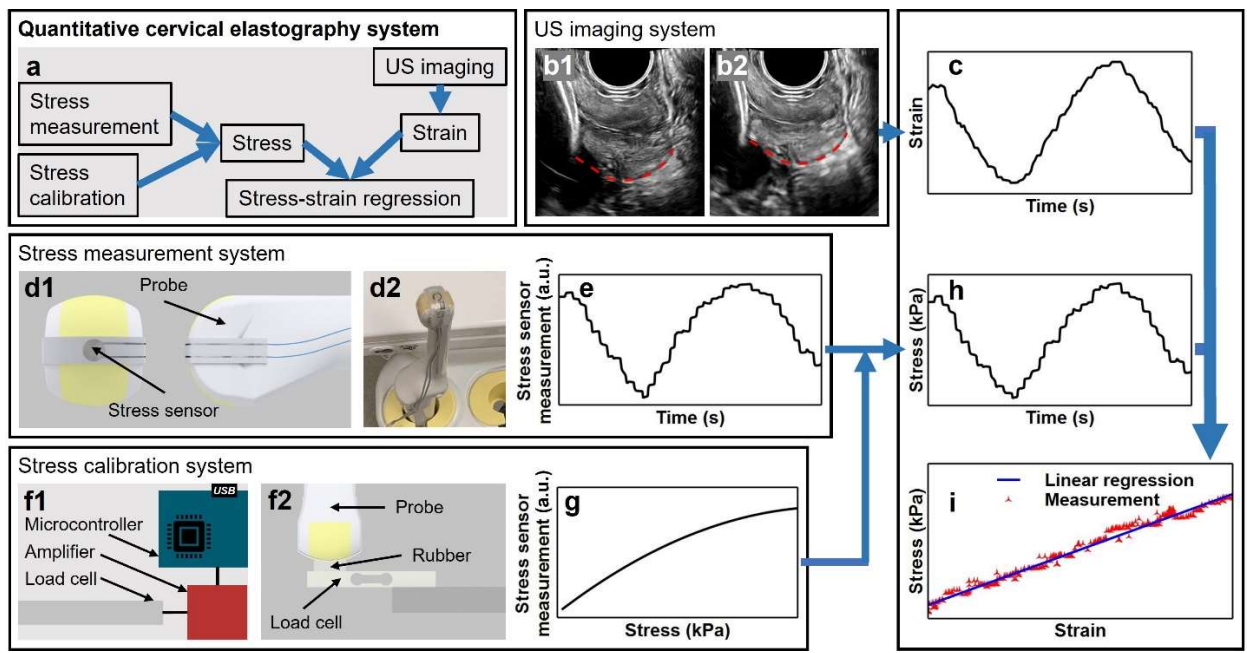

**Figure 1** Quantitative cervical elastography system. **a** The general workflow of the quantitative cervical elastography system. **b1-b2** Two B-mode images of a cervix (posterior boundary marked as a red-dashed curve), showing deformation of the cervix. **c** Strains of the cervix quantified from the continuously acquired B-mode images. **d1-d2** A stress sensor embedded into a layer of silicone sealant and mounted to the probe. **e** A series of measured stresses. **f1-f2** A stress sensor calibration system, consisting of a load cell, a load cell amplifier, and a microcontroller board. During calibration, a layer of silicone rubber is placed between the probe and the load cell to stabilize

contact. **g** A calibration function describing the relation between stress sensor measurement and the true stress. **h** A series of true stresses obtained by calibrating the stress sensor measurements in **e** using the calibration function in **g**. **i** A stress-strain scatter plot and its linear regression using the strains in **c** and the stresses in **h**.

**Accuracy of strain quantification**

Automatization of our quantitative cervical elastography system relies on automatic strain quantification. We demonstrate the accuracy of strain quantification in phantom and participant experiments. In the phantom experiment, we track the thickness of a layer of gelatin phantom while it is vertically pressed by the ultrasound probe. We place a layer of gelatin phantom on a metal plate and use the probe to press the phantom and record the B-mode video. Two frames of the video are shown in **Figure 2a1** and **a2**, respectively. In each frame, the lower boundary of the phantom is marked by a yellow bar. Depths of the lower boundary in different frames are shown in **Figure 2b**, in which, the two black-dotted vertical lines indicate the times of the two frames, respectively, the tracked depths are indicated by red dots, and the direct measurements of the thickness (ground-truth depths) are marked by a blue-solid curve. The root-mean-square (RMS) error between measurements from the two methods is 0.087 mm, which is much smaller than the deformation amplitude of 2.9 mm (error rate: 0.087/2.9 = 3.0%). We further quantify strains $\sigma_{auto}$ and $\sigma_{gt}$ from the automatically tracked depths and the ground-truth depths, respectively. We compare the strains in **Figure 2c**, where the values of ($\sigma_{gt}$, $\sigma_{auto}$) are marked by red dots and the identity relation is indicated by a blue-solid line. The RMS error between $\sigma_{gt}$ and $\sigma_{auto}$ is 0.0019 (3.0% of the amplitude of $\sigma_{gt}$: 0.0642), which shows a high accuracy of the automatic strain quantification algorithm.

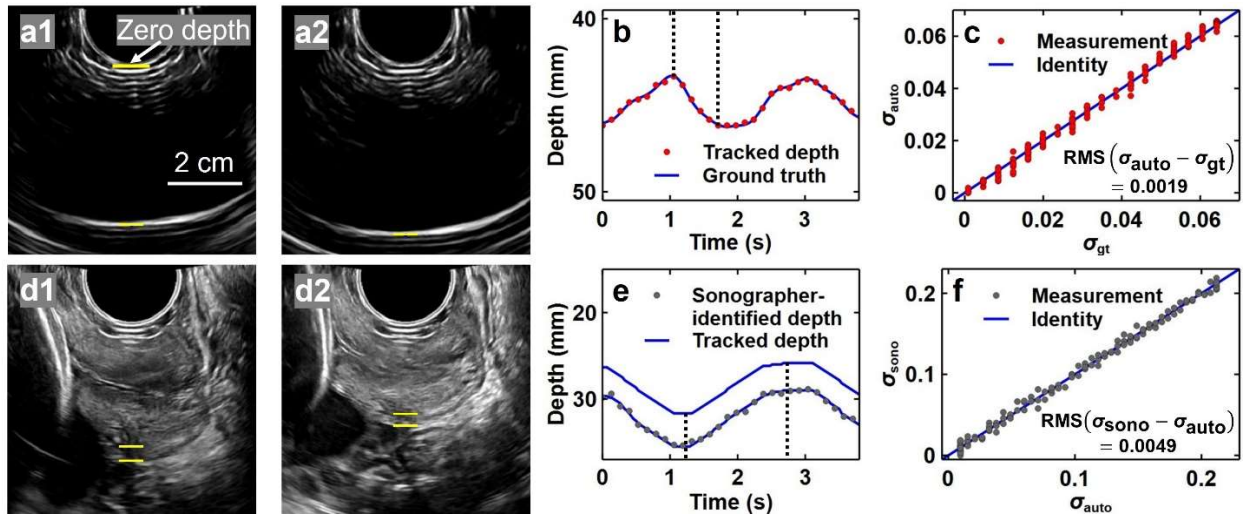

**Figure 2** Correlation-based automatic strain quantification. **a1-a2** Two frames in a B-mode video of a gelatin phantom. The lower boundary of the phantom is marked as a yellow bar. **b-c** Comparison of ground-truth depths and tracked depths of the lower boundary of the phantom (**b**) and the quantified strains (**c**). **d1-d2** Two frames in a B-mode video of a cervix. Two features with center depths marked as yellow bars are tracked. **e-f** Tracked depths of the two features,

sonographer-identified depths of the deeper feature (**e**), and quantified strains of the deeper feature (**f**).

Next, we apply this algorithm to a B-mode video of a participant's cervix. Two frames of the video are shown in **Figure 2d1** and **d2**, respectively. Two features (marked as yellow bars) are tracked, and the depths of the two features in different frames are shown in **Figure 2e**, in which, the times of the two frames are indicated by two black-dotted vertical lines, respectively, and the tracked depths of the two features are denoted by two blue-solid curves. For further validation, we let a sonographer identify the depths of the deeper tracked feature (posterior boundary of the cervix) in all the 120 frames, shown as gray dots in **Figure 2e**. We quantify strains $\sigma_{auto}$ and $\sigma_{sono}$ from the automatically tracked depths and sonographer-identified depths, respectively, and plot in **Figure 2f** the values of ($\sigma_{auto}, \sigma_{sono}$) (gray dots) with an identity (blue-solid) line. The RMS error between $\sigma_{auto}$ and $\sigma_{sono}$ (0.0049) is 2.3% of the amplitude of $\sigma_{auto}$ (0.2118), which further demonstrates the high accuracy of the strain quantification algorithm.

**Accuracy, repeatability, and reproducibility of the system**

We demonstrate the accuracy, repeatability, and reproducibility of the quantitative elastography system in phantoms experiments and a pilot study of 19 pregnant participants. We first visualize the measurement quality using a phantom experiment as an example. In **Figure 3a**, the first frame of a B-mode video of a gelatin phantom is shown. The depths of multiple features of the image are marked as white bars. We apply a threshold of one third of the maximum pixel value to all images in the video to reject weak scattering regions. In the remaining regions, using the correlation-based automatic feature tracking algorithm, we obtain the depths of the marked features in different frames. For each tracked feature, the PCC between the tracked depths and the minus stress sensor measurements is further calculated. Depths and PCCs for the tracked features (close to the lower boundary) are shown in **Figure 3b** and **c**, respectively. The zero-depth layer (probe surface) is marked by a red line and a red dot in **Figure 3b** and **c**, respectively; while the tracked features are marked by blue curves and dots, respectively. The mean depth of the tracked features close to the lower boundary is an estimation of the depth of the lower boundary. The depths of the zero-depth layer and the lower boundary are used to calculate the strains, which are in a range from 0 to 0.111 as shown in **Figure 3d**. The stress calibration of this measurement is shown in **Figure 3e**. The stress sensor measurements and true stresses for the calibration are shown as a scatter plot, while the corresponding nonparametric regression[48] is shown as a blue-solid curve ($f_{cal}$). In this experiment, the stress sensor measurements of interest are in a range from 561 to 760, corresponding to true stresses from 1.25 kPa to 4.06 kPa. In this range of the calibration function $f_{cal}$, the mean difference of the 95% confidence curves (**Supplementary Figure 2c**) is 3.07 (error rate: 3.07/(760−561) ≈ 1.5%), which is negligible for this research. Through calibration, stress sensor measurements for the phantom are transformed into true stresses, as shown in **Figure 3f**. The stresses and strains of this gelatin phantom have a strong linear relation (with $R^2 = 0.9869$), as shown in **Figure 3g**, which agrees with the results in Hall *et al.*[49]. The slope of this linear model is an estimation of Young's modulus of this phantom.

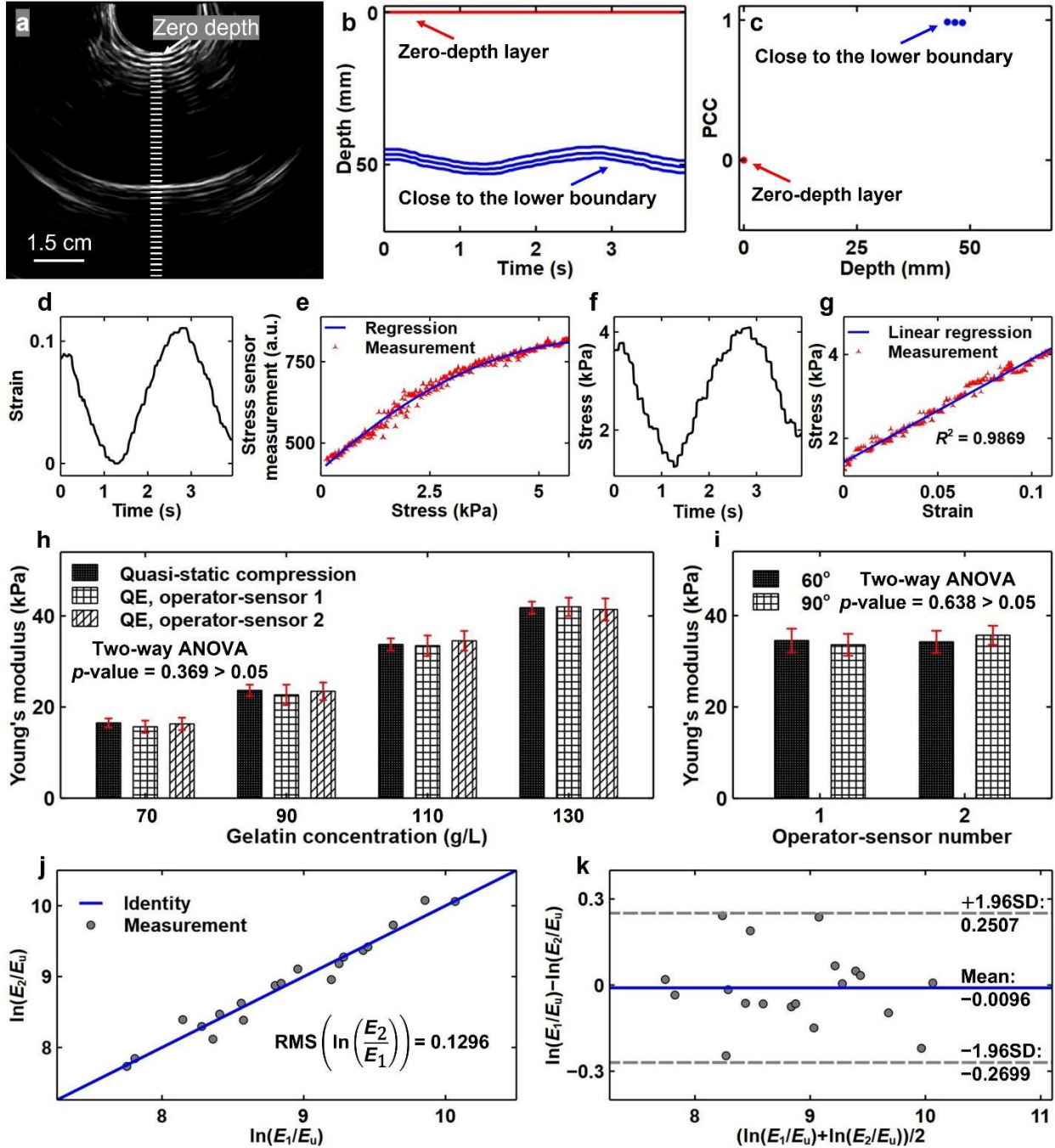

**Figure 3** Accuracy, repeatability, and reproducibility of the quantitative elastography system. **a** The first frame of a B-mode video of a gelatin phantom with depths of multiple features marked as white bars. **b** Depths tracked by the correlation-based algorithm of the features close to the probe surface or lower boundary of the phantom. **c** PCCs between the tracked depths in **b** and the minus stress sensor measurements. **d** Strains calculated between depths of the zero-depth layer and the lower boundary. **e** Calibration of a stress sensor. The scatter plot represents the calibration measurements, and the blue-solid curve is the nonparametric regression. **f** True stresses in one measurement of the phantom. **g** Scatter plot of the stresses and strains with the linear regression.

**h** Comparison of measurements from the quantitative elastography (QE) by two operators using two randomly chosen stress sensor replicas and from the quasi-static compression method for gelatin phantoms of concentrations 70 g/L, 90 g/L, 110 g/L, and 130 g/L, respectively. **i** Measurements of a phantom from QE by two operators using two stress sensors with contact angles of 60° and 90°. **j** Measurements of the 19 participants by two sonographers. Each gray dot of coordinates $\left(\ln\left(\frac{E_1}{E_u}\right), \ln\left(\frac{E_2}{E_u}\right)\right)$ represents a participant with measurements $E_1$ and $E_2$ from the two sonographers, respectively. **k** The Bland-Altman plot of the measurements from the two sonographers.

Next, we apply the proposed quantitative elastography and the standard quasi-static compression method (**Supplementary Figure 4a**) to four gelatin phantoms with concentrations of 70 g/L, 90 g/L, 110 g/L, and 130 g/L, respectively. Two randomly chosen stress sensor replicas are used by two operators, respectively, in the quantitative elastography, and each stress sensor is used to measure each phantom 10 times. The quasi-static compression method is also used to measure each phantom 10 times. As shown in **Figure 3h**, the two operators have similar measurements, and the two methods have similar measurements. Statistically, we test the quantitative elastography's dependency on operator-sensor choice and the method's difference from the quasi-static compression method by performing the two-way ANOVA on these measurements. The two-way ANOVA shows no significant differences ($p$-value = 0.369 > 0.05) among measurements using the quantitative elastography with operator 1 and stress sensor 1 (operator-sensor 1), the quantitative elastography with operator 2 and stress sensor 2 (operator-sensor 2), and the quasi-static compression method. Here, the operator-sensor difference is considered the random block effect. Moreover, we calculate the CVs of these measurements: 0.0854, 0.0987, 0.0672, and 0.0478 (0.0833, 0.0820, 0.0625, and 0.0586) for the four phantoms, respectively, using the quantitative elastography with operator-sensor 1 (operator-sensor 2); 0.0587, 0.0528, 0.0392, and 0.0318 for the four phantoms, respectively, using the quasi-static compression method. The insignificance of the differences in the two-way ANOVA validates the accuracy and reproducibility of the quantitative elastography system and the low values of CV demonstrate the repeatability.

Furthermore, to test contact angles' effects on measurements, we use two stress sensor replicas to measure a gelatin phantom (with a concentration of 110 g/L) with contact angles of 60° and 90° (**Supplementary Figure 4b** and **c**, respectively). As shown in **Figure 3i**, different stress sensors and different contact angles lead to similar results. Quantitatively, the two-way ANOVA shows no significant difference ($p$-value = 0.638 > 0.05) between measurements with contact angles of 60° and 90°. Still, the stress sensor difference here is considered as the random block effect. Based on the test, the proposed system has high robustness with respect to the contact angle, which is important for clinical applications.

Then we test the repeatability and reproducibility of the proposed system in a pilot study of 19 pregnant participants. Two sonographers performed the quantitative elastography at least three times (satisfying the measurement quality requirement) on each participant. The mean CVs of the measurements from the two sonographers are 0.0403 and 0.0540, respectively, showing high repeatability of the system in participant experiments. To simplify the analysis of reproducibility, we denote the geometric means of the two sonographers' measurements of each participant as $E_1$ and $E_2$, respectively. Values of $E_1$ and $E_2$ for the 19 participants are compared in **Figure 3j**: each gray dot of coordinates $\left(\ln\left(\frac{E_1}{E_u}\right), \ln\left(\frac{E_2}{E_u}\right)\right)$ represents a participant with measurements $E_1$ and $E_2$ from the two sonographers, respectively. Here, we define $E_u = 1$ Pa. All these gray dots are close to the identity line with an error quantified by the RMS of $\ln\left(\frac{E_2}{E_u}\right) - \ln\left(\frac{E_1}{E_u}\right) = \ln\left(\frac{E_2}{E_1}\right)$: 0.1296. This value corresponds to a mean relative difference of $e^{0.1296} - 1 \approx 13.84\%$ between the two sonographers' measurements $E_1$ and $E_2$. The PCC between $\ln\left(\frac{E_1}{E_u}\right)$ and $\ln\left(\frac{E_2}{E_u}\right)$ is 0.981, showing high reproducibility between operators. Furthermore, we compare the measurements in a Bland-Altman plot (**Figure 3k**), which shows that the measurement differences are within the mean ± 1.96SD (standard deviation) range and further validates the agreement between the sonographers' measurements.

**Longitudinal imaging of pregnant participants**

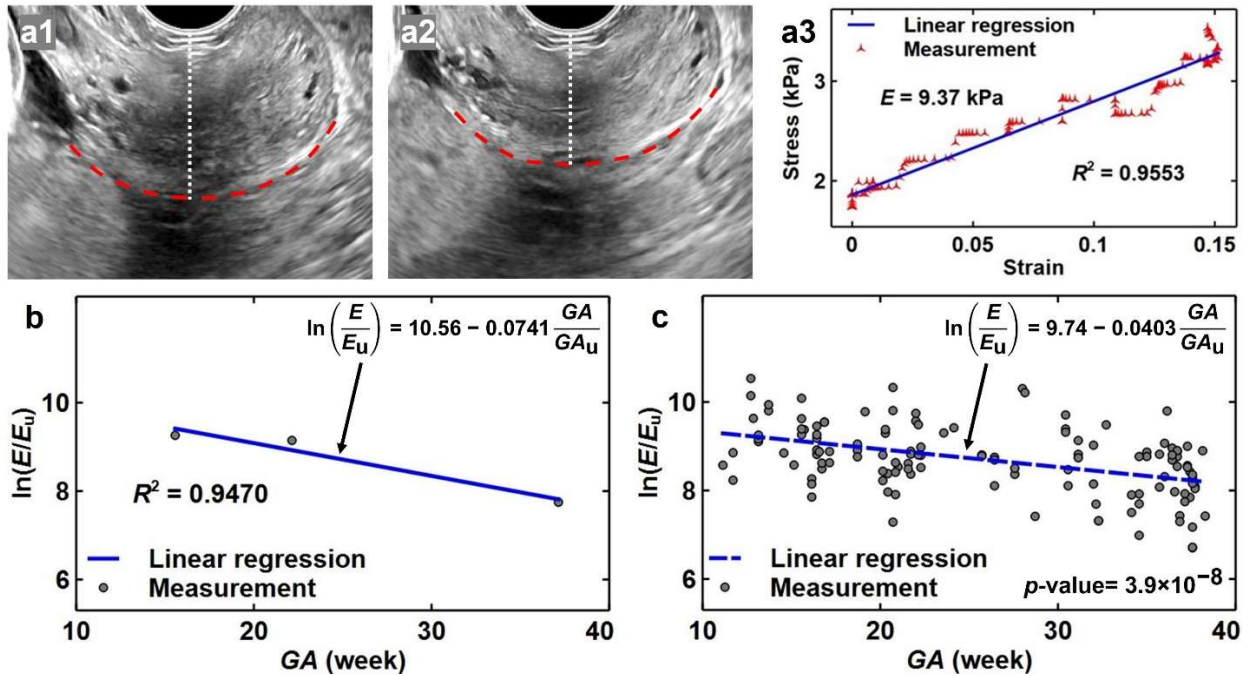

**Figure 4** Longitudinal imaging of pregnant participants. **a1-a2** Transvaginal cervical B-mode images (with different stresses) of a participant in one video. The posterior boundary of the cervix is marked by a red-dashed curve. The cervical tissue thickness is indicated by a white-dotted line. **a3** Measured stress-strain values (red three-pointed stars) of this participant in this B-mode video and their linear regression (blue-solid line). A Young's modulus of 9.37 kPa is estimated in the regression. **b** All three measurements (gray dots), expressed as $\ln\left(\frac{E}{E_u}\right)$, of a participant and their linear regression (blue-solid line). **c** Measurements (gray dots), expressed as $\ln\left(\frac{E}{E_u}\right)$, of all participants and their linear-mixed-model regression (blue-dashed line).

We apply the quantitative cervical elastography system in pregnant participants to quantify cervical Young's moduli longitudinally over pregnancy. The overall success rate of the measurements (the fraction of the measurements that meet the measurement quality requirement) is 54.7%. Performing multiple measurements in every imaging session guarantees a high imaging-session success rate. For example, assuming that different measurements are independent, 3 (4) measurements yield an imaging-session success rate of 90.7% (95.8%). One measurement of a participant is shown in **Figure 4a1–a3**. B-mode images of the cervix in the sagittal plane are shown first with minimal stress applied to the tissue (**Figure 4a1**) where the cervical tissue is noted to be thicker and then as stress is applied (**Figure 4a2**) the cervical tissue is noted to be thinner. In each image, the posterior boundary of the cervix is marked by a red-dashed curve while the cervical tissue thickness is indicated by a white-dotted line. Strains quantified from the B-mode images are plotted against the measured stresses in **Figure 4a3** (red three-pointed stars). A linear regression (blue-solid line) is applied to estimate the Young's modulus $E$ = 9.37 kPa with $R^2 = 0.9553$.

Quantitative measurement of cervical elasticity allows for the comparison of values from a single participant in different imaging sessions. We analyze three measurements of a participant in three visits, expressed as $\ln\left(\frac{E}{E_u}\right)$ and denoted by three gray dots, respectively, in **Figure 4b**. Applying linear regression to the three measurements, we obtain $\ln\left(\frac{E}{E_u}\right) = 10.56 - 0.0741\frac{GA}{GA_u}$ (with $R^2 = 0.9470$), denoted as a blue-solid line in **Figure 4b**. Here, $GA$ denotes gestational age and we define $GA_u = 1$ week. This linear regression is equivalent to the exponential decaying of shear wave speed during pregnancy[8,33]. The linear regression result means that the initial cervical Young's modulus of this participant was approximately $e^{10.56}E_u \approx 38.56$ kPa. During pregnancy, the cervical Young's modulus of this participant decreased by approximately $1 - e^{-0.0741} \approx 7.14\%$ per week.

Quantitative cervical elastography also allows for the comparison of elasticity among participants. For each participant with three or more visits, we perform linear regression to $\ln\left(\frac{E}{E_u}\right)$ and obtain $\ln\left(\frac{E}{E_u}\right) = \beta_0 - \beta_1\frac{GA}{GA_u}$. The values of $\beta_1$ for these participants have a mean value of 0.0476 with an SD of 0.0302, indicating a general trend of cervical softening and individual variation. Further, we apply a linear-mixed-model regression to $\ln\left(\frac{E}{E_u}\right)$ of all measurements of these participants and obtain the population average softening trend as $\ln\left(\frac{E}{E_u}\right) = 9.74 - 0.0403\frac{GA}{GA_u}$. Values of $\ln\left(\frac{E}{E_u}\right)$ and the linear regression are shown as gray dots and a blue-dashed line, respectively, in **Figure 4c**. We detect significant softening of the cervix as gestational age increases ($p$-value < 0.001). Specifically, we quantify that the geometric mean value of the initial cervical Young's modulus is approximately $e^{9.74}E_u \approx 16.98$ kPa, and during pregnancy the cervical Young's modulus decreases by $1 - e^{-0.0403} \approx 3.95\%$ per week. Here, the geometric mean value of Young's modulus $E$ corresponds to the arithmetic mean value of $\ln\left(\frac{E}{E_u}\right)$. Based on the linear regression, for the first ($11 \leq GA \leq 13$), second ($14 \leq GA \leq 26$), and third ($27 \leq$ $GA \leq 40$) trimesters, the geometric mean values of the cervical Young's moduli are 13.07 kPa, 7.59 kPa, and 4.40 kPa, respectively.

**DISCUSSION**

We developed a quantitative cervical elastography system based on transvaginal ultrasound imaging and stress measurement that can be applied to patients during pregnancy to quantify the cervical softening process. Specifically, this system overcomes limitations of prior systems by simultaneously measuring stress and tissue strain, providing a quantitative, operator-independent assessment of cervical tissue elasticity which can be used in the same patient over pregnancy and can be used for between-patient comparisons (development of population norms) as well as within-patient comparisons (identifying patterns of normal or abnormal cervical remodeling in an individual). Importantly, this system is based on minimal modifications to a transvaginal ultrasound probe which is routinely used during pregnancy for other clinical indications and thus is safe, familiar to patients, familiar to healthcare providers, and widely available in obstetric ultrasound units.

The results of the phantom experiments with known elasticities demonstrate accurate and robust agreement among repeated measurements of our system even when the angle of stress application is varied, which addresses the reality of use *in vivo* where the angle of imaging may vary from patient to patient. Furthermore, the findings from the longitudinal study in pregnant patients demonstrate that the cervix softens over pregnancy, matching the expected pregnancy physiology and previous research.

Technological advancements include the GUI which gives the sonographer real-time information on measurement quality. Additionally, the automatic strain quantification through correlation-based feature tracking enables real-time data analysis and visualization. Although initial stress is still required for good contact, this system captures the stress-strain curve for a large range of stress and applies a linear regression to the curve to estimate Young's modulus, which effectively minimizes the measurement's dependency on initial stress. Importantly, it does not rely on a homogeneous-medium assumption, which is likely violated given the anatomic characteristics of the cervix. Moreover, the stress measurement system and strain quantification algorithm are independent of ultrasound machines, which means that they can be interfaced with different machines with relatively minor software updates. This cross-platform feature allows these measurements to be adapted regardless of the ultrasound platform used, making the technology scalable clinically and commercially.

In summary, quantification of cervical tissue elasticity will allow investigations documenting normal and abnormal cervical softening patterns in pregnancy and can be used in clinical populations to quantify risks for obstetric disease associated with abnormal cervical physiology (preterm birth, post-term birth, cervical insufficiency, etc.). This technology has wide applicability to research endeavors and multiple potential clinical uses.

Future updates to improve the system include improving the precision of estimating Young's moduli in anatomic region-specific areas of cervical tissue, which could begin to map different geographic changes within the cervical anatomy over pregnancy. For example, Young's modulus specifically of the anterior cervical region, posterior cervical region, or area most proximal to the uterine cavity could be interrogated individually to ascertain global versus region-specific softening and association with obstetric disease. Continued optimization of the stress calibration system, ease of sensor application, use of multiple sensors and finite element mechanical modeling, and/or automatic image processing are all steps that could be refined to make this technology more efficient and scalable in the future.

## DISCLOSURES

Lihong V. Wang has financial interests in Microphotoacoustics, Inc., CalPACT, LLC, and Union Photoacoustic Technologies, Ltd., which did not support this work.

## ACKNOWLEDGEMENTS

We thank Emily Diveley, Stephanie Pizzella, and Cassandra Hardy for their help in the study. This project was supported by the March of Dimes Prematurity Research Center (3125-17303A).

# Supplementary materials

**Supplementary Note 1 Graphical user interface (GUI)**

We developed a GUI using Python for the quantitative elastography system to provide centralized data acquisition (B-mode video recording and stress measurement) and to give real-time feedback to sonographers (**Supplementary Figure 1a**). The GUI allows for real-time streaming of B-mode video from the ultrasound machine to the computer. During measurements, the acquired data are visualized in two windows: one for B-mode video, the other for stress sensor measurement.

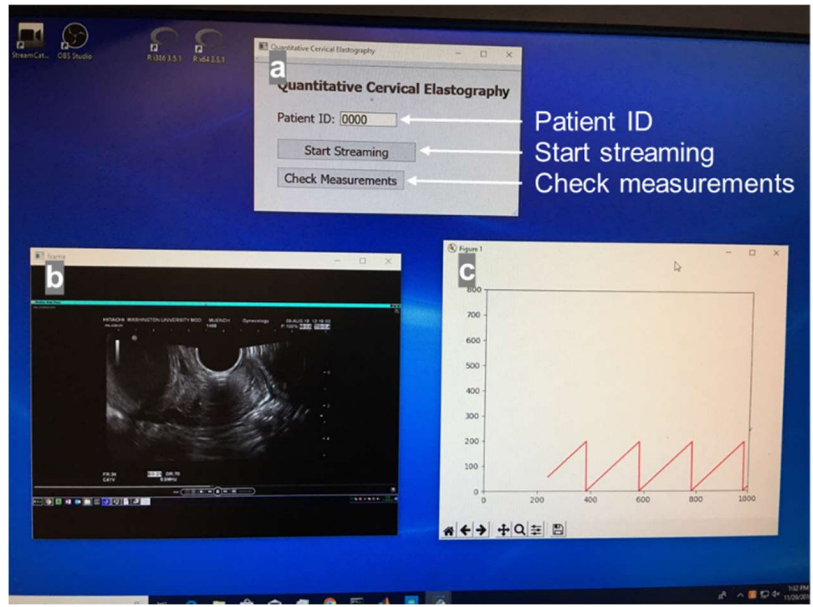

**Supplementary Figure 1** GUI for the quantitative elastography system. **a** Control panel. **b** A window for B-mode video streaming. **c** A window for stress sensor measurement visualization.

In an imaging session, a sonographer first puts in the participant ID and clicks the "Start Streaming" button to start B-mode video streaming. Then the sonographer adjusts the probe location and orientation to achieve high image quality and applies press-release cycles to the cervix using the probe. Stress sensor measurements require a small amount of memory and are all recorded and saved, whereas saving B-mode video for the whole imaging session requires a large amount of memory and may not be necessary. Thus, B-mode video (120 frames) only in the latest 4 seconds is cached in the computer's random-access memory (RAM). For example, the $n$-th frame is stored in the RAM for the $(\text{rem}(n-1, 120) + 1)$-th image. Here $\text{rem}(p, q)$ denotes the remainder after the division of an integer $p$ by another integer $q$. After performing 2 to 4 press-and-release cycles, the sonographer clicks the "Check Measurements" button to save all the stress sensor measurements and the latest recorded 120 frames (automatically reorganized to their true order) of the B-mode video for analysis. As a demonstration, a video played on another computer is streamed in **Supplementary Figure 1b**, while numerically generated stress sensor measurements are shown in **Supplementary Figure 1c**.

**Supplementary Note 2 Stress calibration**

In theory, the stress calibration is a 1D regression process. We denote a series of load cell measurements in calibration as $m_l, l = 1,2,...,L_{cal}$, which correspond to stress sensor measurements $\hat{\sigma}_l, l = 1,2,...,L_{cal}$. Here, $L_{cal}$ denotes the number of sampling points in the calibration. From the load cell measurements, we obtain the true stresses

$$\sigma_l = \frac{m_l g}{A}, l = 1,2,...,L_{cal}, \tag{1}$$

where $g = 9.81 \text{ m} \cdot \text{s}^{-2}$ is the gravitational acceleration and $A$ is the contact area. Load cell measurements are relatively accurate, whereas the stress sensor measurements are noisy. To reduce noises in the stress sensor measurements, we perform median filtering of kernel size three (with a duration of 0.0375 s) to $\hat{\sigma}_l$. A press-release cycle has a duration of one to three seconds; thus, the median filtering has negligible effects on the general trend of stress variation while reducing noise locally. Then $\hat{\sigma}_l$ and $\sigma_l$ are sorted in the ascending order of $\sigma_l$, resulting in $\hat{\sigma}_{l'_l}$ and $\sigma_{l'_l}$, $l = 1,2,...,L_{cal}$, respectively, where $(l'_1, l'_2,...,l'_{L_{cal}})$ is a permute of $\{1,2,...,L_{cal}\}$. To further reduce noises in the stress sensor measurements, we apply median filtering of kernel size three to $\hat{\sigma}_{l'_l}$. Performing a nonparametric regression[1] to $\hat{\sigma}_{l'_l}$ and $\sigma_{l'_l}$, we obtain the calibration function

$$\hat{\sigma}_{l'_l} \approx f_{cal}(\sigma_{l'_l}), l = 1,2,...,L_{cal}. \tag{2}$$

The accuracy of this calibration is estimated from the 95% confidence lower and upper curves in this nonparametric regression. After the calibration, we transform any stress sensor measurement $\tilde{\sigma}$ to true stresses $f_{cal}^{-1}(\tilde{\sigma})$ by inverting Equation (2).

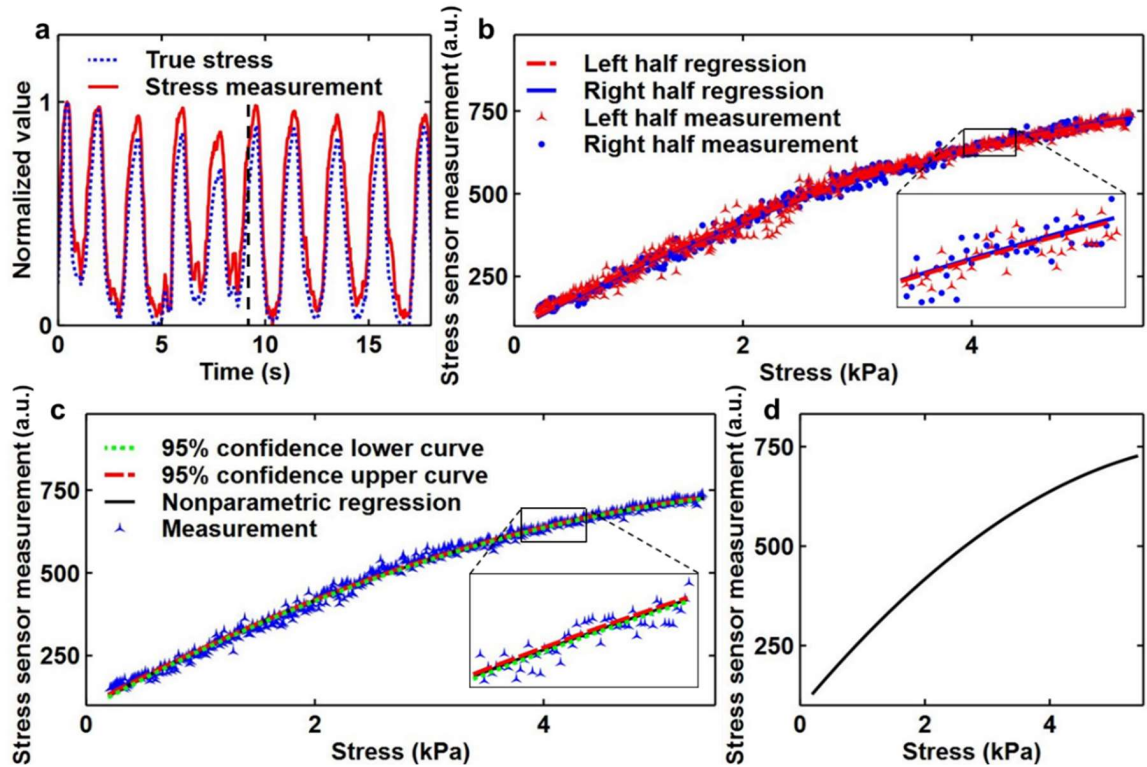

**Supplementary Figure 2** Stress calibration in one experiment. **a** Normalized true stresses (measured by a load cell, marked as a blue-dotted curve) and normalized stress sensor

measurements (red-solid curve) in 18 seconds. A black-dashed line is marked at 9 seconds. **b** Scatter plots of stress sensor measurements and true stresses for the left half (0–9 seconds, red three-pointed stars) and right half (9–18 seconds, blue dots), respectively, and their nonparametric regressions (red-dashed curve and blue-solid curve, respectively). **c** Scatter plot (blue three-pointed stars) of the true stresses and stress sensor measurements in the whole 18 seconds, and their nonparametric regression (black-solid curve) with the 95% confidence lower (green-dotted curve) and upper (red-dashed curve) curves. **d** The final calibration function.

We demonstrate the stress calibration through one experiment, in which the normalized true stresses and normalized stress sensor measurements are shown as a blue-dotted curve and a red-solid curve, respectively, in **Supplementary Figure 2a**. A black-dashed line at 9 seconds divides the true stresses and stress sensor measurements into two parts: the left half (0–9 seconds) and the right half (9–18 seconds). Scatter plots of the stress sensor measurements and true stresses for the left half and right half are shown as red three-pointed stars and blue dots, respectively, in **Supplementary Figure 2b**. The corresponding nonparametric regressions[1] are shown as a red-dashed curve and a blue-solid curve, respectively. For this experiment, the stress sensor measurements of interest are between 310 and 625. In this range, the mean difference between the two regression curves is 2.37 (error rate: 2.37/(625−310) = 0.75%), which is negligible for this research and shows the high inner consistency of the stress sensor. A closeup subset in the black-boxed region in **Supplementary Figure 2b** is used to show the details.

In practice, the true stresses and stress sensor measurements (blue three-pointed stars) in the whole 18 seconds are used for nonparametric regression[1] (black-solid curve), as shown in **Supplementary Figure 2c**. The 95% confidence lower (formed by lower values of the 95% confidence intervals) and upper (formed by upper values of the 95% confidence intervals) curves are shown as a green-dotted curve and a red-dashed curve, respectively, in **Supplementary Figure 2c**. The mean difference between the two 95% confidence curves is 3.48 (error rate: 3.48/(625−310) = 1.1%), which is negligible in this research. A closeup subset in the black-boxed region is used to show the details. The final calibration function in **Supplementary Figure 2d** shows the relation between the true stress and the stress sensor measurement.

**Supplementary Note 3 Automatic feature tracking**

A critical component of the quantitative cervical elastography system is automatic strain quantification from continuously acquired B-mode images, which relies on feature tracking across these images. Specifically, the posterior boundary of the cervix can be identified in the first image and its depth can be tracked in other images. The cervical thicknesses in different images can be obtained by subtracting the depths of the posterior boundary and the probe surface. Strains then can be calculated from the thicknesses. To automatically compute the strain of cervical tissue due to applied stress, we develop a correlation-based automatic feature tracking algorithm.

We perform automatic feature tracking in the frames of a continuously acquired B-mode video. We construct a coordinate system in each frame of the video using the probe tip as the origin, the downward vertical direction as the $z$-axis, and the rightward horizontal direction as the $x$-axis. Then we express the B-mode video as $f_l(x, z), l = 1, 2, \ldots L$, where $L$ denotes the number of frames. In this research, stresses on the cervical tissue are applied along the $z$-axis, causing cervical deformation in the same direction. Therefore, the tracking algorithm is performed only along the $z$-axis. For convenience, we denote $g_l(z) = f_l(0, z)$. In the first frame, we choose features (values on vertical line segments) of length $2z_f$ and centered at depths $z_{1,n} = \frac{Z(n-1)}{N}, n = 1,2, \ldots, N$ along the $z$-axis. Here, $Z$ represents the largest depth of interest. We further denote the depths of the centers of these features in the $l$-th frame as $z_{l,n}, l = 2, \ldots, L, n = 1,2, \ldots N$. Due to spatiotemporal continuity of cervical deformation, a feature has similar values in different frames, which is mathematically expressed as

$$g_l(z + z_{l,n}) \approx g_{l'}(z + z_{l',n}), z \in [-z_f, z_f], l, l' \in \{1,2, \ldots, L\}, n = 1,2, \ldots N. \quad (3)$$

Based on Equation (3), we propose a correlation-based algorithm to track the $n$-th feature's movement from the $l$-th frame to the $l'$-th frame:

$$\hat{z}_{l',n} = \underset{z' \in [0,Z]}{\mathrm{argmax}}\, \mathrm{PCC}_{z \in [-z_f, z_f]}\left(g_l(z + z_{l,n}), g_{l'}(z + z')\right), l, l' \in \{1,2, \ldots, L\}, n = 1,2, \ldots N. \quad (4)$$

Here, we estimate the $n$-th feature's depth in the $l'$ frame ($\hat{z}_{l',n}$) based on the feature's depth in the $l$-th frame ($z_{l,n}$) using the Pearson correlation coefficient (PCC) of two functions in an interval:

$$\mathrm{PCC}_{z \in [z_1, z_2]}(p(z), q(z)) =$$

$$\frac{\int_{z_1}^{z_2} \left(p(z) - \frac{1}{z_2 - z_1}\int_{z_1}^{z_2} p(z')dz'\right)\left(q(z) - \frac{1}{z_2 - z_1}\int_{z_1}^{z_2} q(z')dz'\right)dz}{\sqrt{\int_{z_1}^{z_2} \left(p(z) - \frac{1}{z_2 - z_1}\int_{z_1}^{z_2} p(z')dz'\right)^2 dz}\sqrt{\int_{z_1}^{z_2} \left(q(z) - \frac{1}{z_2 - z_1}\int_{z_1}^{z_2} q(z')dz'\right)^2 dz}}. \quad (5)$$

We can directly obtain $\hat{z}_{l',n}$ for all $l' = 2, \ldots, L$ and $n = 1,2, \ldots N$ by letting $l = 1$ and solving the optimization problems described by Equation (4). However, this direct algorithm is not robust in practice. Due to accumulative deformation, $|z_{l',n} - z_{1,n}|$ can be large, meaning $z_{l',n}$ needs to be searched in a large range, which is not robust due to similarities between different features and large deformations of a single feature. To increase the robustness in the search of $z_{l,n}$, we reduce Equation (4) to

$$\hat{z}_{l,n} = \underset{z' \in [\hat{z}_{l-1,n} - z_s, \hat{z}_{l-1,n} + z_s]}{\mathrm{argmax}}\, \mathrm{PCC}_{z \in [-z_f, z_f]}\left(g_{l-1}(z + \hat{z}_{l-1,n}), g_l(z + z')\right),$$
$$l = 2,3, \ldots, L, n = 1,2, \ldots N. \quad (6)$$

Here, $z_s$ denotes the half search length, and we recursively obtain $\hat{z}_{l,n}, l = 2,3, \ldots, L$ from $\hat{z}_{1,n}$ (defined as $z_{1,n}$) based on Equation (6). Due to small deformation between adjacent frames (30-Hz frame rate), the deformation of a single feature is negligible and a small value is assigned to $z_s$, which means that the recursive search of $z_{l,n}$ based on Equation (6) is more robust than the direct search based on Equation (4). In practice, to reduce the error caused by tissue movement along the $x$-axis, we apply spatial lowpass filtering along the $x$-axis to all images $f_l(x, z), l = 1,2, \ldots, L$ before the feature tracking. Because we analyze the strain only along the probe axis in this research, the feature tracking (the most time-consuming part of data processing) is performed only along the probe axis, which takes a few seconds and is negligible in practice.

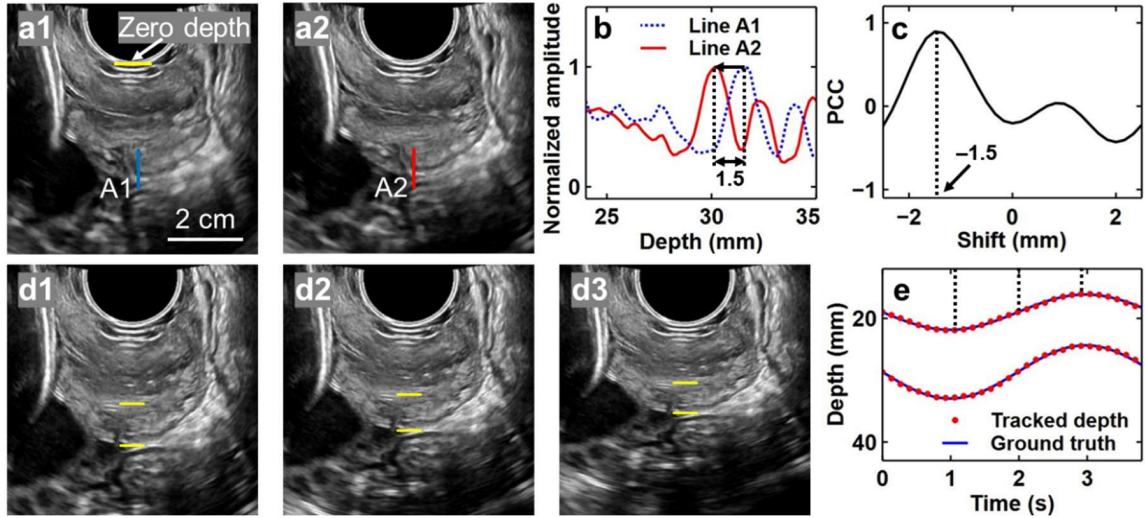

**Supplementary Figure 3** Correlation-based automatic feature tracking. **a1**-**a2** Two B-mode images showing the deformation of a cervix. The probe surface is marked as zero depth (a yellow bar). Two vertical lines A1 and A2 with the same length are picked from the images at the same tissue region (cervical posterior region) right below the probe and are marked as a blue line and a red line, respectively. **b** Comparison of the values on lines A1 and A2, marked as a blue-dotted curve and a red-solid curve, respectively. **c** PCCs between the values on line A1 (with different shifts) and the values on line A2. **d1**–**d3** Three frames in a video simulated from a truly acquired B-mode image and predefined strains. The tissue is tracked at two features (center depths marked as yellow bars) using the correlation-based method. **e** Comparison of the ground-truth depths of the two features and the results from the correlation-based automatic feature tracking algorithm.

We first use two B-mode images of a cervix (shown in **Supplementary Figure 3a1** and **a2**, respectively) to demonstrate the correlation-based automatic feature tracking algorithm. We choose two vertical lines at the same tissue region (cervical posterior region) in both images for analysis, marked as a blue line A1 and a red line A2 in **Supplementary Figure 3a1** and **a2**, respectively. Values on lines A1 and A2 are shown in **Supplementary Figure 3b** as a blue-dotted curve and a red-solid curve, respectively. Here, we set the depth of the probe surface as zero, as indicated by a yellow bar in **Supplementary Figure 3a1**, and define the depth of a feature (values on a vertical line segment) as its center's vertical distance to the probe surface. We calculate the PCC between the values on two segments of the same length from the two lines with different vertical shifts. Performing an ergodic search of local vertical shift to line A1, from –2.5 mm to +2.5 mm, we obtain a series of correlation coefficients. The shift-dependent correlation coefficients are shown in **Supplementary Figure 3c**. As we see, with line A1 shifted vertically by −1.5 mm, values on these two lines have the best match, meaning that from **Supplementary Figure 3a1** to **a2**, features on line A1 move 1.5 mm closer to the zero depth. Therefore, the optimal shift is defined as where the maximum PCC is achieved.

Then we validate the feature-tracking algorithm's performance on a video simulated from a truly acquired B-mode image and predefined strains. Given a B-mode image $f(x, z)$, we define its deformation by strain $\epsilon$ as

$$f_\epsilon(x,z) = f\left(xe^{-\frac{\epsilon}{2}}, ze^\epsilon\right). \tag{7}$$

Based on this definition and a series of strains $\epsilon_l, l = 1,2, \ldots, L$, we obtain a series of B-mode images $f_{\epsilon_l}(x,z) = f\left(xe^{-\frac{\epsilon_l}{2}}, ze^{\epsilon_l}\right), l = 1,2, \ldots, L$. Three representative frames of the video are shown in **Supplementary Figure 3d1–d3**. Two features are tracked by the correlation-based algorithm. The depths of the features' centers are marked by two yellow bars for convenience of comparison. The tracked depths of the two features are shown in **Supplementary Figure 3e** as red dots, while ground-truth depths (calculated with predefined strains) of the two features are plotted as blue-solid curves. Times of the three chosen frames are indicated by three black-dotted vertical lines, respectively. Root-mean-square error of the tracked depths of these two features compared to the ground-truth depths are 0.12 mm and 0.14 mm, respectively, which are much smaller than the corresponding deformation amplitudes 5.7 mm (error rate: $0.12/5.7 \approx 2.1\%$) and 8.6 mm (error rate: $0.14/8.6 \approx 1.6\%$). The correlation-based automatic feature tracking algorithm shows high accuracy in this numerical simulation.

**Supplementary Note 4 Periodic correction**

The recursive feature tracking based on Equation (6) generates an accumulative error in $\hat{z}_{l,n}$. To reduce the error, we let each sonographer periodically press and release the cervix using the probe and acquire B-mode videos with durations longer than one period. After the feature tracking based on Equation (6), a period starting with the first frame and ending with the $L'$-th frame is identified. Due to the periodicity, deformation between the first frame and the $L'$-th frame is small, which allows us to use Equation (4) with $l' = 1$ and $l = L'$ to find a corrected version of $\hat{z}_{L',n}$ (denoted as $\hat{z}'_{L',n}$) robustly in a small search range. Based on the correction to the $L'$-th frame, we update other frames in this period through linear interpolation:

$$\hat{z}'_{l,n} = \hat{z}_{l,n} + \frac{l-1}{L'-1}\left(\hat{z}'_{L',n} - \hat{z}_{L',n}\right), l = 1,2, \ldots, L', n = 1,2, \ldots N. \tag{8}$$

Frames out of this period ($L' < l \leq L$) are corrected using frames (with the same phases) in this period based on Equation (4) with a small search range. We call this process the periodic correction.

**Supplementary Note 5 Strain quantification**

From the tracked feature depths $\hat{z}'_{l,n}$, we obtain the logarithmic strain of the tissue between the $n$-th and $n'$-th ($n' < n$) features in the $l$-th frame

$$\epsilon_{n',n,l} = \ln\frac{z_{1,n} - z_{1,n'}}{\hat{z}'_{l,n} - \hat{z}'_{l,n'}}, l = 1,2, \ldots, L. \tag{9}$$

In this research, we analyze the average strain of the cervix cross section ($n' = 1, n = n_p$):

$$\epsilon_l = \epsilon_{1,n_p,l} = \ln\frac{z_{1,n_p} - z_{1,1}}{\hat{z}'_{l,n_p} - \hat{z}'_{l,1}}, l = 1,2, \ldots, L. \tag{10}$$

Here, $n_p$ denotes the index of the feature indicating the posterior boundary of the cervix.

**Supplementary Note 6 Phantom experiment**

In the standard quasi-static compression method to measure Young's modulus, a gelatin phantom is sandwiched between two disks, as shown in **Supplementary Figure 4a**. The lower disk is placed on a scale while the upper disk is fixed to a linear stage through a cylinder. The linear stage is used to move the upper disk to compress the phantom and record the compression distance. The scale is used to measure the compression force. The recorded distance and force are used to calculate the strain and stress, from which, Young's modulus is obtained.

For the proposed quantitative elastography, we test contact angles of 60º and 90º. The case for the contact angle of 60º is shown in **Supplementary Figure 4b**. The black-dotted line indicates the axis of the probe, which forms a 60º contact angle with the phantom surface. The blue-dashed horizontal line is in the phantom surface while the blue-dashed vertical line is normal to the surface. The three lines are in the same plane. The black-dotted arrow (parallel to the probe axis) indicates the moving direction of the probe. The case for contact angle of 90º is shown in **Supplementary Figure 4c**. The blue-dashed vertical line indicates the axis of the probe, which is normal to the phantom surface. The blue-dashed arrow (parallel to the probe axis) indicates the moving direction of the probe.

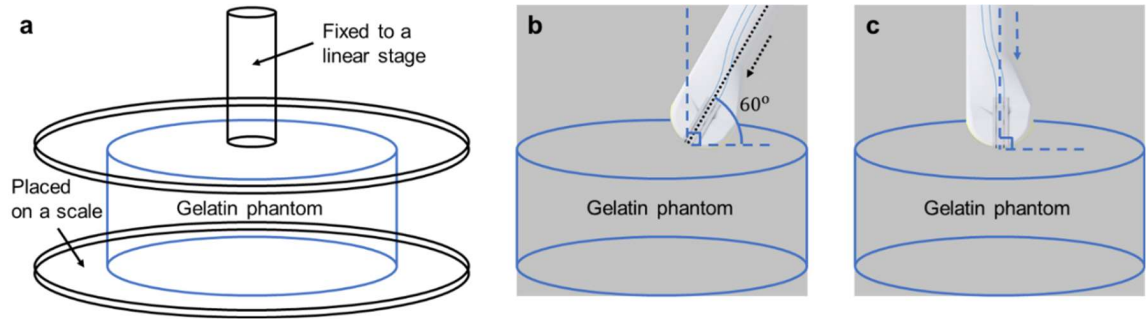

**Supplementary Figure 4** Using the standard quasi-static compression method and the proposed quantitative elastography to measure Young's modulus of a gelatin phantom. **a** Using the standard method to measure Young's modulus of a gelatin phantom. **b-c** Using the quantitative elastography to measure Young's modulus of a phantom with contact angles 60º and 90º.

**Supplementary Table 1** Characteristics of the 22 participants included in this study

| Characteristic | Value |
|---|---|
| Number of participants | 22 |
| GA (weeks, mean (SD)) | 38.87 (1.05) |
| BMI (kg/m², mean (SD)) | 26.09 (7.77) |
| Maternal Age (years, mean (SD)) | 28.37 (9.00) |
| Race (%) | |
|    African-American | 9 (40.9) |
|    White | 11 (50.0) |
|    Others | 2 (9.1) |
| Preterm birth (%) | |
|    No | 21 (95.5) |
|    Yes | 1 (4.5) |
| Nulliparous (%) | |
| No | 11 (50.0) |
| Yes | 11 (50.0) |
| Education (%) |  |
| Unknown | 4 (18.2) |
| Less than high school | 1 (4.6) |
| Completed high school | 5 (22.7) |
| College graduate | 5 (22.7) |
| Advanced degree | 7 (31.8) |
| Smoke (%) |  |
| No | 19 (86.4) |
| Yes | 3 (13.6) |
| Alcohol (%) |  |
| No | 22 (100.0) |
| Yes | 0 (0.0) |

GA, gestational age; SD, standard deviation; BMI, body mass index.